\documentclass[]{aa}

\usepackage{txfonts}
\usepackage{graphicx}
\usepackage{natbib}
\usepackage{rotating}
\usepackage{amssymb}

\bibpunct{(}{)}{;}{a}{}{,} 

\def\mpc2{\ {\rm h}_{70} \ M_{\sun}.{\rm pc}^{-2}}

\begin{document}

\title{A CFH12k Lensing Survey of X-ray Luminous Galaxy Clusters}
\subtitle{I: Weak Lensing Methodology}

\titlerunning{CFH12k Weak Lensing Survey -- Abell 1689}

\author{S. Bardeau\inst{1}
  \and
  J.-P. Kneib\inst{1,2}
  \and
  O. Czoske\inst{3,1}
  \and
  G. Soucail\inst{1}
  \and
  I. Smail\inst{4}
  \and
  H. Ebeling\inst{5}
  \and
  G. P. Smith\inst{2}
}

\offprints{S. Bardeau}

\institute{
  Observatoire Midi-Pyr\'en\'ees, UMR5572,
  14 Avenue Edouard Belin,
  31400 Toulouse, France.
  \and
  Caltech, Astronomy, 105-24, Pasadena, CA 91125, USA.
  \and
  Institut f\"ur Astrophysik und Extraterrestrische
  Forschung, Auf dem H\"ugel 71, 53121 Bonn, Germany.
  \and
  Institute for Computational Cosmology, University of Durham,
  South Road, Durham DH1 3LE, UK.
  \and
  Institute for Astronomy, University of Hawaii, 2680 Woodlawn Dr,
  Honolulu, HI 96822, USA
}

\date{Received ---; accepted ---}

\abstract{We describe the weak lensing methodology we have applied to
  multi-colour CFH12k imaging of a homogeneously-selected sample of
  luminous X-ray clusters at $z\sim 0.2$. The aim of our survey is to 
  understand the variation in cluster structure and dark matter profile
  within rich clusters. The method we describe converts a fully reduced
  CFH12k image into constraints on the cluster mass distribution in two 
  steps: (1) determination of the ``true'' shape of faint (lensed) 
  galaxies, including object detection, point spread function (PSF)
  determination, galaxy shape measurement with errors; (2) conversion of 
  the faint galaxy catalogue into reliable mass constraints using a range
  of 1D and 2D lensing techniques. Mass estimates are derived independently
  from each of the three images taken in separate filters to quantify the
  systematic uncertainties. Finally, we compare the cluster mass model to
  the light distribution of cluster members as derived from our imaging 
  data. To illustrate the method, we apply it to the well-studied cluster
  \object{Abell 1689} ($z=0.184$). In this cluster, we detect the 
  gravitational shear signal out to $\sim 3$\,Mpc at $>$3-$\sigma$
  significance. The two-dimensional mass reconstruction has a
  $\sim$10-$\sigma$ significance peak centered on the brightest cluster
  galaxy.  The weak lensing profile is well fitted by a NFW mass profile
  with $M_{200} = 14.1^{+6.3}_{-4.7} \times 10^{14}\,M_{\sun}$, and
  $c = 3.5^{+0.5}_{-0.3}$ ($\chi^2=0.33$), or by a power law profile with
  $q=0.75\pm0.07$ and $\theta_{\rm E} = 14\farcs6\pm0\farcs3$
  ($\chi^2=0.64$). The mass-to-light ratio is found to be almost constant
  with radius with a mean value of $M/L_R = 150\,h\,(M/L)_{\sun}$.
  We compare these results to other weak lensing analyses of Abell~1689 
  from the literature and find good agreements in terms of the shear 
  measurement as well as the final mass estimate.
   }

\maketitle

\section{Introduction}

\begin{figure*}
  \centering
  \includegraphics[width=\textwidth]{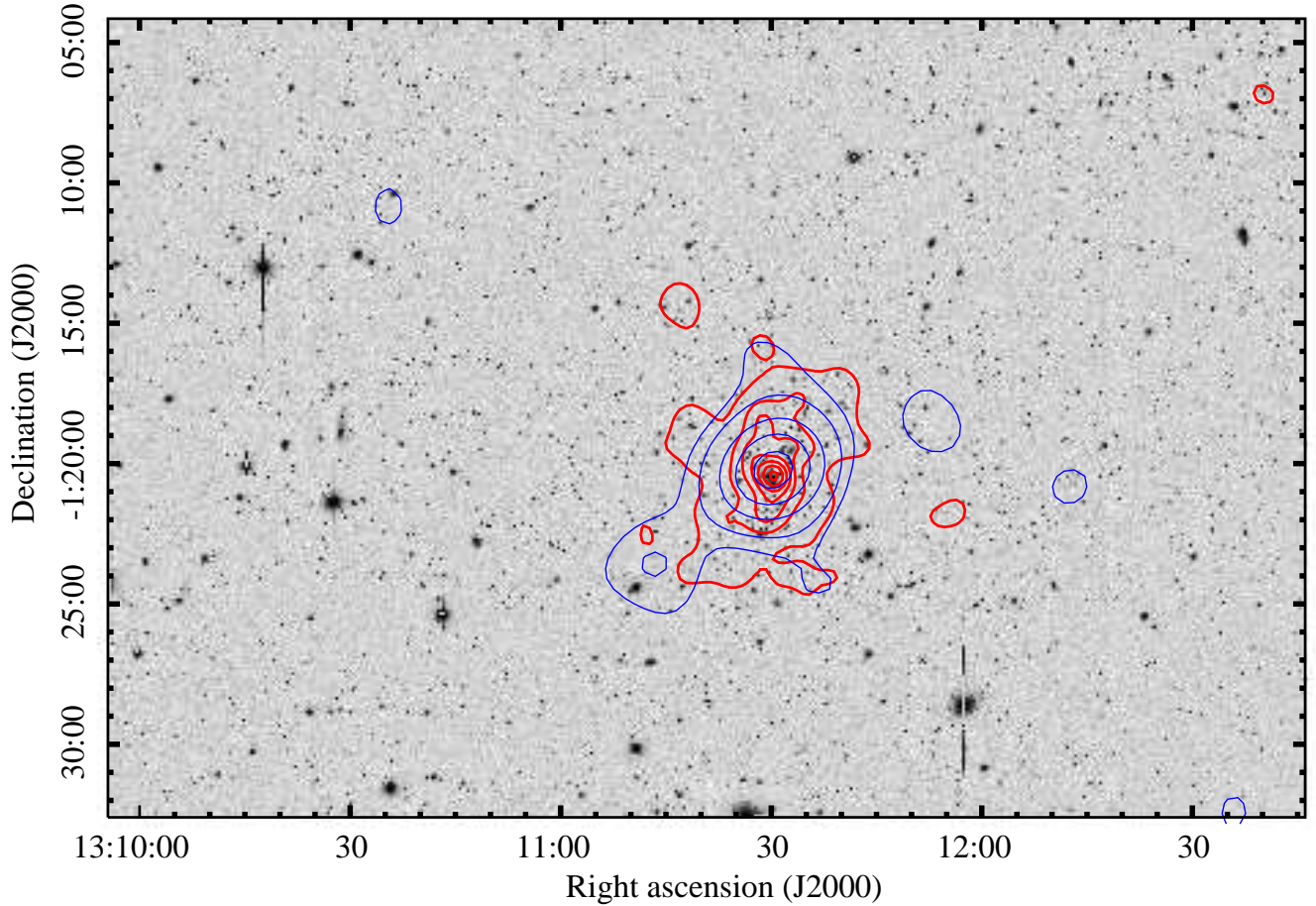}
  \caption{The full $42\arcmin \times 28 \arcmin$ CFH12k R-band image
    of Abell 1689. The thick contours represent the number density of
    bright galaxies selected in the R-band: the first contour corresponds
    to 5 objects per square arcmin, increased by steps of 5 units. The 
    maximal density is 36.5 galaxies ${\rm arcmin}^{-2}$ in the cluster 
    center. The thin contours represent the mass density reconstruction 
    with \textsc{LensEnt2} and an ICF of 180\arcsec\ (see 
    Sect.~\ref{ssec:lensenttxt} for more details). Contour levels are 
    respectively 2, 3, 5, 7 and $9\sigma$, while the peak value corresponds
    to a mass density of $1100 \mpc2$.  North is to the top, East to the
    right.}
  \label{fig:image}
\end{figure*}

Clusters of galaxies are the most massive collapsed structures 
in the Universe.  They are located
at the nodes of the filamentary cosmic web, as mapped by 
the  SDSS and 2dF redshift surveys.  These massive systems are the focus of
both theoretical \citep[e.g.\ ][]{eke96,bahcall97,viana98} and
observational studies. The aim is to better understand cluster
formation and evolution and thus it is important to quantify their
physical properties as precisely
as possible (e.g.\ mass distribution, mass density profile,
importance of substructure, etc.).  Different techniques such as
galaxy dynamics, X-ray emission, the Sunyaev-Zeldovich effect or gravitational
lensing, are available to probe the physical properties of clusters.
Gravitational lensing is a particularly attractive method as it is
directly sensitive to the total mass distribution irrespective of its
physical state \citep[see the review by][]{Mellier1999}.

 Although the study of a single cluster can be instructive, we need to
study homogeneous samples of massive clusters in order to better understand
cluster physics, test theoretical predictions and to constrain the 
cosmological and physical parameters governing the growth of structure in 
the Universe. Indeed, clusters are expected to show some variation in their
properties, in particular in the amount of substructure and their merger 
history, which can be directly probed by measuring their mass distribution.
Thus to obtain a representative view of the properties of clusters, a fair 
and statistically-reliable sample of cluster needs to be studied.

 In order to obtain a better understanding of the mass distributions on 
small and large scales in clusters, we have selected a sample of 11 X-ray 
luminous clusters \citep{Czoske2003,Smith2005} identified in the XBACs 
sample \citep[X-ray Brightest Abell-type Clusters:][]{ebeling96}. All these
clusters have X-ray luminosities of $L_{\rm X}\ge 8 \times 
10^{44}\,\mathrm{erg\,s^{-1}}$ in the 0.1--2.4 keV band, and all lie in
a narrow redshift slice at $z\sim 0.2$ (from $z_{\rm A2218} = 0.171$ to 
$z_{\rm A1835} = 0.253$). As XBACS is restricted to Abell clusters 
\citep{abell89}, it is X-ray flux-limited but not truly X-ray selected.
However, a comparison with the X-ray selected {\it ROSAT} Brightest Cluster
Sample \citep[BCS:][]{ebeling98,ebeling00} shows that $\sim 75\%$ of the 
BCS clusters in the redshift and X-ray luminosity range of our sample are 
in fact Abell clusters. Hence, our XBACs sample is, in all practical
aspects, indistinguishable from an X-ray selected sample.

 Using the CFH12k wide field camera \citep{Cuillandre2000} mounted at
the Canada-France-Hawaii Telescope (CFHT), we imaged all 11 clusters
in our sample in the B, R and I bands. In the present paper we
present the {\it weak lensing methodology} we have applied to analyse
these images, using our observations of Abell~1689 as a test case.\\

 The first step of any weak lensing work is to correct the observed galaxy
ellipticities for any observational smearing: circularization or
anisotropy due to the point spread function (PSF). The classical approach to
do this is the so-called KSB method \citep{kaiser95}, implemented in the
\textsc{imcat} software \citep[see also][]{Luppino-Kaiser1997,rhodes00,
Kaiser2000}. The basic idea is to relate the ``true'' ellipticity of the
background sources to the observed ellipticity through polarizability
tensors, which include the smearing effect of the PSF, possibly with
anisotropic components. In practice these can be computed through the
combination of the second order moments of the light distribution of the
galaxies and the PSF itself. 
However, in this paper we will use an {\it inverse} approach through a maximum
likelihood or Bayesian estimate of the source galaxy shape convolved by
the local PSF \citep[this method was first proposed by][]{kuijken99}.
Both the galaxy shape and the local PSF are
modeled in terms of sums of elliptical Gaussians. This approach is
implemented in the software \textsc{Im2Shape} which has been developed
by \citet{bridle01}.  The main advantage of \textsc{Im2Shape} is that
it provides estimates of the uncertainties of the recovered
parameters of the sources and these uncertainties can then be included
in the mass inversion. 

In the weak lensing limit, the ellipticities of background
galaxies give an unbiased estimate of the shear field induced by the
gravitational potential of the foreground cluster. The estimate is
inherently noisy due to the shape measurement errors and
the intrinsic ellipticities of the galaxies. 
Several methods have been proposed to reconstruct the mass density
field (or the potential) of the foreground structure from the
measured shear field. Non-parametric methods are usually best to
produce a mass-map, necessary to identify mass peaks. They can also
be used to estimate the cluster mass profile by means of the
aperture mass densitometry method \citep{fahlman94,schneider96}.
On the other hand parametric methods are best to constrain the cluster
mass profile and total mass by fitting a radial shear profile to the
galaxy ellipticities.

 To illustrate the various methods and techniques used, we apply our
procedure to one particulary well-studied cluster from our sample, 
Abell~1689. Abell~1689 at $z=0.184$ is  one of the richest clusters ($R=4$)
in the Abell catalog. Abell~1689 is a powerful cluster lens and has been 
studied by various groups using different lensing techniques 
\citep{Tyson1990, Tyson-Fischer1995, Taylor1998, clowe01, King2002}. It has
also been studied in X-rays using {\sl Chandra} \citep{Xue-Wu2002} and
{\sl XMM-Newton} \citep{Andersson-Madejski2004}. The central
structure of this cluster is complex: from the redshift distribution
of 66 cluster members \citet{Girardi1997} find evidence for a
superposition of several groups along the line of sight to the cluster
center which explains the extraordinarily high velocity dispersion of
$2355^{+238}_{-183}\,\mathrm{km\,s^{-1}}$. \citet{czoske04} has
recently obtained a new large dataset of more than 500 cluster galaxy
redshifts in this cluster, which will help elucidate the galaxy
distribution Abell~1689. Preliminary analysis of these data shows
that the large scale distribution of galaxies in and around Abell~1689
is in fact rather smooth and that significant substructure seems
confined to the very center of the cluster. Thus, even though
the cluster has clear substructure, it may still be reasonable 
to model the large-scale mass distribution of the cluster with simple
models, such as the ``universal'' mass profile proposed by
\citet{Navarro1997} (NFW).

This paper is organized as follows: \S\ref{sec:observations}
briefly presents the observations of Abell~1689 used in this paper and
gives a summary of the data reduction procedure and the conversion of
the reduced data into catalogues that can be used in the weak lensing
analysis. In \S\ref{sec:psftxt} we present the measurement of
galaxy shapes and correction for PSF anisotropy using
\textsc{Im2Shape}. In \S\ref{sec:shear_measurements} we convert
the galaxy shape measurements into two-dimensional shear maps and
radial shear profiles.  \S\ref{sec:modelling} explains how we
model the lensing data using both 1D and 2D techniques. 
In \S\ref{sec:light-distribution} we compare the projected mass to the light
distribution in the cluster. Finally in \S\ref{sec:conclusions} we discuss our
method and results.  In a separate paper (Bardeau et al.\ 2005, in
prep.)  we will present a more extensive analysis
of the mass distribution in Abell~1689
combining weak and strong lensing mass measurements.

 We assume $H_0 = 70\,\mathrm{km\,s^{-1}\,Mpc^{-1}}$, $\Omega_{\rm m}=0.3$,
$\Omega_\Lambda=0.7$. At $z=0.18$, $1\arcsec$ corresponds to 
$3.09\,\mathrm{kpc}$ (and $1\arcmin$ to $185\,\mathrm{kpc}$).

\section{Observations and Cataloging}
\label{sec:observations}

We observed Abell~1689 with the CFH12k camera through the B, R and I
filters (Fig.~\ref{fig:image} shows the R-band image) between 30 May
and 2 June 2000.  The camera consists of 12 CCD chips of
$2\mathrm{k}\times 4\mathrm{k}$ pixels with a total field of view of
$42\arcmin \times 28\arcmin$ at a pixel scale of $0\farcs205$.  The
log of the observations of Abell~1689 ($\alpha_{\rm
  J2000}\!=\!13^{\rm h}11^{\rm m}30^{\rm s}$, $\delta_{\rm
  J2000}\!=\!-01\degr20\arcmin28\arcsec$) is summarized in the first
part of Table~\ref{tab:a1689params}.

\begin{table*}
  \caption{Observing log for Abell~1689. We indicate the number of 
           detections in each filter (B, R or I), their number density 
           (expressed in arcmin$^{-2}$, in parenthesis), and the magnitude 
           cuts for galaxy classification. Estimated average redshifts 
           $\bar{z}$ and $\bar{\beta} = \langle D_{\rm ls}/D_{\rm s} 
           \rangle$, with their standard deviations, are given for the 
           faint galaxy catalogues (see \S\ref{ssec:meanzgal} for 
           details).}
  \centering
  \begin{tabular}{lrrrrrrrrr}
\hline
\noalign{\medskip}
\large{Filter}       & \multicolumn{3}{c}{\large{B}}          & \multicolumn{3}{c}{\large{R}}          & \multicolumn{3}{c}{\large{I}}          \\
\noalign{\smallskip}
\hline
\noalign{\smallskip}
Date of observation  & \multicolumn{3}{c}{May 30/June 2 2000} & \multicolumn{3}{c}{May 30/June 2 2000} & \multicolumn{3}{c}{May 30/June 2 2000} \\
Number of exposures  & \multicolumn{3}{c}{4}                  & \multicolumn{3}{c}{5}                  & \multicolumn{3}{c}{5}                  \\
Exposure time (sec)  & \multicolumn{3}{c}{3600}               & \multicolumn{3}{c}{3000}               & \multicolumn{3}{c}{3000}               \\
Seeing               & \multicolumn{3}{c}{$0.91\arcsec$}      & \multicolumn{3}{c}{$0.85\arcsec$}      & \multicolumn{3}{c}{$0.88\arcsec$}      \\
Completeness mag     & \multicolumn{3}{c}{24.9}               & \multicolumn{3}{c}{24.3}               & \multicolumn{3}{c}{22.6}               \\
PSF anisotropy       & \multicolumn{3}{c}{$0.032\pm0.012$}    & \multicolumn{3}{c}{$0.071\pm0.019$}    & \multicolumn{3}{c}{$0.064\pm0.028$}    \\
\noalign{\smallskip}
\hline
\noalign{\smallskip}
Number of Detections &                       & 34669 & (28.6) &                       & 41067 & (33.9) &                       & 28805 & (23.7) \\
\ Stars              &                       &  2223 &  (1.8) &                       &  3488 &  (2.9) &                       &  2397 &  (2.0) \\
\ Galaxies           &                       & 25823 & (21.3) &                       & 30189 & (24.9) &                       & 21145 & (17.4) \\
\ Others             &                       &  6623 &  (5.5) &                       &  7390 &  (6.1) &                       &  5263 &  (4.3) \\
\ \ Bright galaxies  &              B$<$22.0 &  1171 &  (1.0) &              R$<$21.1 &  2166 &  (1.8) &              I$<$19.3 &   950 &  (0.8) \\
\ \ Faint galaxies   &       22.5$<$B$<$25.4 & 20186 & (16.7) &       21.6$<$R$<$24.7 & 22794 & (18.8) &       19.8$<$I$<$23.3 & 14382 & (11.8) \\
\ \ Other galaxies   &       22.0$<$B$<$22.5 &       &        &       21.1$<$R$<$21.6 &       &        &       19.3$<$I$<$19.8 &       &        \\
                     &           or B$>$25.4 &  4466 &  (3.7) &           or R$>$24.7 &  5229 &  (4.3) &           or I$>$23.3 &  5813 &  (4.8) \\
\noalign{\smallskip}
\hline
\noalign{\smallskip}
Faint galaxies $\bar{z}$     & \multicolumn{3}{c}{1.02$\pm 0.42$} & \multicolumn{3}{c}{1.06$\pm 0.42$} & \multicolumn{3}{c}{0.82$\pm 0.35$}     \\
Faint galaxies $\bar{\beta}$ & \multicolumn{3}{c}{0.70$\pm 0.08$} & \multicolumn{3}{c}{0.69$\pm 0.08$} & \multicolumn{3}{c}{0.65$\pm 0.07$}     \\
\noalign{\smallskip}
\hline
\end{tabular}

  \label{tab:a1689params}
\end{table*}

\subsection{Data reduction}
\label{ssec:data-reduction}

 For a detailed description of the data reduction see 
\citet{Czoske2002diss}. Here we just give a brief outline. Pre-reduction of
the CFH12k data was done in a standard way using the 
\textsc{iraf}\footnote{IRAF is distributed by the National Optical
  Astronomy Observatories, which are operated by the Association of
  Universities for Research in Astronomy, Inc., under cooperative
  agreement with the National Science Foundation.} 
package \textsc{mscred} \citep{Valdes1998} for bias subtraction and
flat-fielding using twilight sky images.

 Fringing in the I band images was removed by subtracting a correction 
image constructed from eight science images from different fields taken 
during the same night, after masking any objects detected in the images. 
The appropriate scaling for the fringe correction was determined 
interactively.

 Weak lensing applications demand precise measurements of the shapes of 
faint galaxies and therefore precise relative astrometric alignment of the
individual dithered exposures of the field ($\sim6\arcsec$ in our case). A
transformation is needed between each chip of the input image and a common 
astrometric output grid which has to account for the position of the chip 
in the focal plane, rotation, variations in the height (and possibly tilt) 
of the chip surface with respect to the focal plane, as well as any optical 
distortion induced by the telescope and camera optics. Fourth order 
polynomials were found to be sufficient to model these effects. The method 
that we have developed follows the approach described by \citet{Kaiser1999}.

We use Digital Sky Survey
(DSS\footnote{\texttt{http://www-gsss.stsci.edu/dss/dss\_home.htm,
    http://cadcwww.dao.nrc.ca/dss/}}) images to define the external
reference frame for observations, but then minimize the RMS dispersion of 
the transformed object coordinates from all the exposures rather than the
deviations between the transformed object coordinates from the 
corresponding DSS coordinates for each individual exposure. This approach 
ensures optimal \emph{relative} alignment of the transformed exposures. The
resulting RMS dispersion of the transformed coordinates is of order 
$0\farcs01$, corresponding to 0.05 of a CFH12k pixel, for usually 
$\gg\! 100$ objects per chip.

 The input images are resampled onto the output grid with pixel size
$0\farcs205$ (the median effective pixel scale of the CFH12k camera) using
the software \textsc{swarp} (Version 1.21). Pixel interpolation uses the 
{\sc lanczos3} kernel which preserves object counts, without introducing 
strong artifacts around image discontinuities \citep{Bertin2001}. Fields 
with a large number of exposures ($\ge 10$) were averaged after rejecting
outliers, those with fewer exposures median combined.

 The images were photometrically calibrated on fields of standard
stars taken from the list of \citet{Landolt1992} with additional
photometry by \citet{Stetson2000}. Atmospheric extinction was
determined from sequences of science images spanning a sufficient
range in airmass to allow accurate determination of the extinction
coefficient.

\subsection{Object detection}
\label{ssec:object detection}

With the reduced and calibrated images in hand, the weak shear
information must be extracted from the photometric catalogues.  The
analysis of the images involves a number of steps that we describe in
detail below. These various steps are controlled in (as much as
possible) an automatic way using different \textsc{perl} scripts which
allow a simple and easy handling of catalogues and can easily call
external programmes.

In the present paper we first treat the images taken in the three
filters B, R and I \emph{independently}. Differences between the
results obtained from the three datasets are expected due to a number
of effects.  Different seeing in the images affects the accuracy of
the measurement of galaxy shapes and hence the accuracy of the derived
shear fields. Different photometric depths of the images will change
the number density of faint background galaxies and thus again the
accuracy of the shear measurements. Finally, the images sample
different wavebands of the observed galaxies, which has an effect on
the contrast between cluster and background galaxies if these are
selected based on magnitude alone.  This independent approach allows
us to assess the uncertainties introduced by the listed effects.
Of course it is desirable to eventually combine the information
present in the three images in an optimal way so as to arrive at
definitive measurements of the physical properties of the cluster.  A
first attempt at this combination is implemented here but will be
discussed in more detail in a forthcoming paper.

The first step is to construct a master photometric catalogue of each
individual image. For this purpose and to automate the procedure as
much as possible we have used \textsc{sextractor}
\citep{Bertin-Arnouts1996} in a two-pass mode.  A first run is made
to detect bright objects, with a detection level of $5\sigma$ above
the background. The average size (full width at half maximum, FWHM) of
the point spread function (PSF) is then easily determined from
  the sizes of stars. The saturation level of the image is also
determined in this run.  These parameters are then fed into a second
\textsc{sextractor} run with a lower detection level ($1.5\,\sigma$
with a minimum size of 5 connected pixels above the threshold). This
second output catalogue corresponds to the working catalogue. The
total number of objects detected in each image is given in
Table~\ref{tab:a1689params}. The photometry was computed using the
\textsc{mag\_auto} method of \textsc{sextractor}.

\subsection{Star catalogue}
\label{ssec:magmutxt}

\begin{figure}
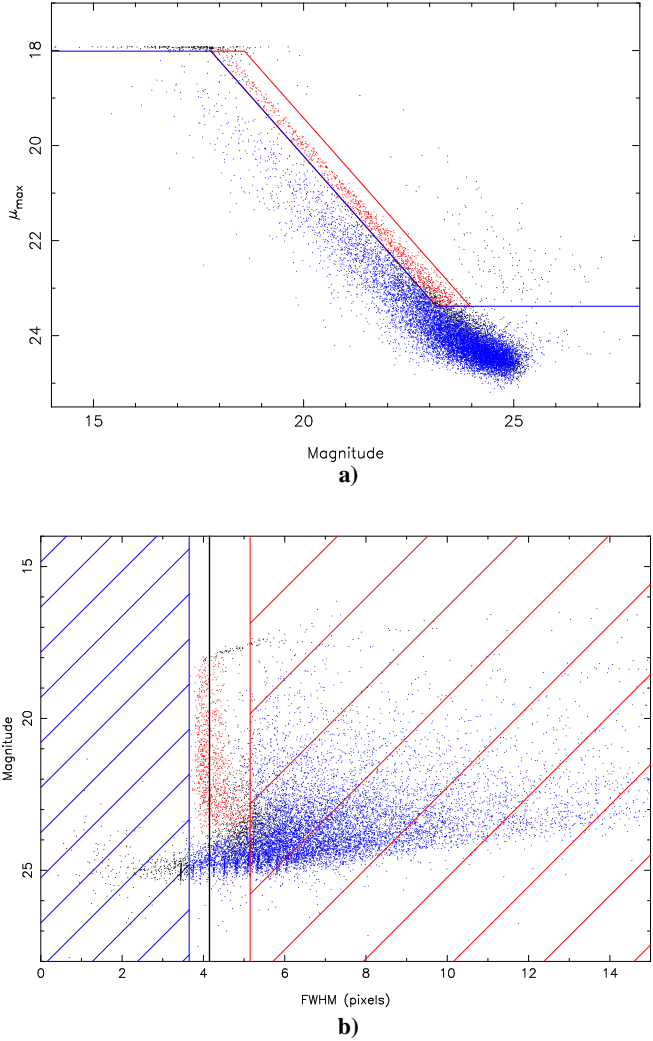

  \centering
  \includegraphics[angle=-90,width=0.48\textwidth]{1643f02a.ps}\\
  \hspace{.5cm} \textbf{a)}\\
  \vspace*{.3cm}
  \includegraphics[angle=-90,width=0.48\textwidth]{1643f02b.ps}\\
  \vspace*{.1cm}
  \hspace{.5cm} \textbf{b)}
  \caption{\textbf{a)} Magnitude-$\mu_{\rm max}$ diagram for all objects
    detected by \textsc{sextractor} in the Abell~1689 R-band image.  The
    points inside the parallelogram correspond to stars, the points
    below to galaxies.  Points on the upper left correspond to cosmic
    rays, defects and saturated objects. \textbf{b)} FWHM-magnitude
    diagram for all the objects detected by \textsc{sextractor} in the
    Abell 1689 R-band image. The vertical black line indicates the
    average seeing value (4.15 pixels for the Abell 1689 R-band
    image).  Stars are excluded from the right hatched part of the
    diagram ($> \mathrm{seeing}+1$ pixel), and galaxies from the left
    hatched part ($< \mathrm{seeing}-0.5$).}
  \label{fig:magmups}
\end{figure}

The second step is to extract a star catalogue from the full catalogue
which will then be used to estimate the local PSF. We select stars by
a number of criteria.  First we locate objects in the magnitude --
$\mu_{\rm max}$ diagram (Fig.~\ref{fig:magmups}a) where $\mu_{\rm
  max}$ is the central surface brightness of the objects.  Stars, for
a given flux, have the highest peak surface brightness (provided
they do not saturate the CCD).  Hence they populate the ``star''-region
of Fig.~\ref{fig:magmups}a, limited to a maximum value of the peak
surface brightness by the saturation of the detector, and to a lower
value, where galaxies start to overlap the star sequence.

We use an additional cut in FWHM indicated on Fig.~\ref{fig:magmups}b:
objects with $\mathrm{FWHM} > \mathrm{seeing} + 1$ pixel are excluded
from the star catalogue. Note that very compact objects (in the
upper-right part of Fig.~\ref{fig:magmups}a) correspond to cosmic rays
or noise defects in the overlapping region between chips. They are
rejected and are put in the ``others'' catalogue (see
Table~\ref{tab:a1689params}).

Finally, the star catalogue is cleaned one last time (see 
\S\ref{ssec:psfmaptxt}) once the star shapes are adequately measured
by \textsc{im2shape}.

\subsection{Galaxy catalogues}
\label{ssec:galscats}

The third step in our
analysis is to compute the galaxy catalogues that will be used
to identify the faint lensed galaxies and the bright galaxies that are
likely to be part of the cluster and which will be used to calculate
the cluster luminosity.

Galaxies are selected from the Magnitude-$\mu_{\rm max}$ diagram (see
Fig.~\ref{fig:magmups}a).  First, as for the stars, saturated
galaxies are excluded. We checked that none of the brightest galaxies
in the cluster core are affected by this cut which only affects lower
redshift galaxies.  Furthermore, we applied two additional cuts:
galaxies must have a \textsc{sextractor} \textsc{class\_star} parameter
lower than 0.8 (this removes faint stars or faint compact galaxies
from the catalogue), and galaxies cannot be smaller than stars, so we
exclude all objects with a FWHM smaller than $\mathrm{seeing}\! -\! 0.5$
pixel. This blind cleaning is done in a similar way in all three bands.
These cuts remove most of the defects in the catalogues.

The galaxy catalogue is then split into three sub-catalogues, defined by
their magnitude range: one for the brightest galaxies, dominated by
the cluster members, one for the faintest galaxies expected to be
background sources, and the last one for the remaining galaxies
(intermediate magnitude range galaxies or excluded objects).

The bright galaxies catalogue is defined with respect to the
apparent $m^*$ of cluster galaxies (see
  \S\ref{ssec:mlprof} for the estimate of $m^*$ in each filter).
In order to achieve good contrast between cluster galaxies and 
the background field population, while still 
integrating a fair fraction of the cluster
luminosity function, we define the bright galaxy catalogue 
by selecting galaxies down to $m^*+2$ for the B and I-band and
$m^*+3$ for  the R-band
(the deeper R-band image allows to have a fainter limit).
For Abell~1689, these correspond to magnitude limits of $B < 22.0$, $R <
21.1$ and $I < 19.3$.
 For illustration, a rough estimate of the field contamination is
given for the Abell~1689 R catalogue: outside a radius $r = 10 \arcmin$ the
galaxy density measured in the magnitude range $R < 21.1$ is 1.3
gal\,arcmin$^{-2}$ while the galaxy density in an inner radius $r = 5
\arcmin$ is 5.5 gal\,arcmin$^{-2}$. Therefore with our selection criteria the
field contamination does not exceed 20 to 25\% of the ``bright galaxy''
catalogue which will be called hereafter the ``cluster catalogue''. After a
uniform correction for field contamination, it will be used to
measure the cluster luminosity and derive a light map,  providing
simple comparisons of the stellar
distributions between clusters in our survey.

\begin{figure}
  \centering
  \includegraphics[angle=-90,width=0.48\textwidth]{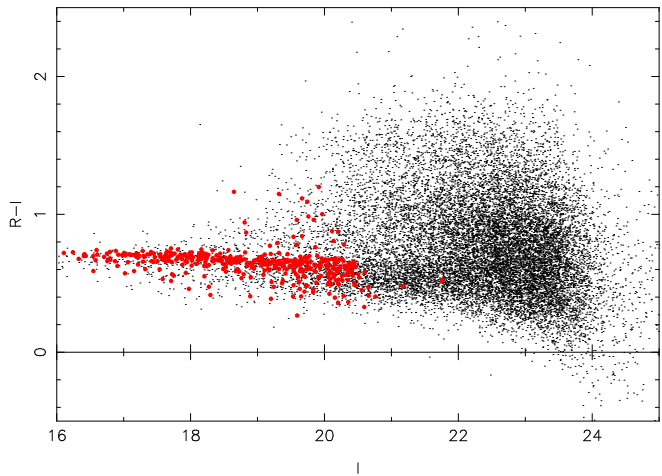}
  \caption{Colour-magnitude diagram for the galaxies detected in the R and I
    filters. Magnitudes are the {\sc mag\_best} measurement from
    \textsc{sextractor}, and colours are computed from magnitudes measured
    in a $3\arcsec$ aperture. Larg (red) points are the \emph{R bright
    galaxies} (as defined in Sect.~\ref{ssec:galscats}) within
    $300\arcsec$ of the cluster centre.}
  \label{fig:R-I_I}
\end{figure}

Fig.~\ref{fig:R-I_I} shows the colour ($R-I$) -- magnitude ($I$) diagram for 
the galaxies matched in both R and I filters
The \emph{red sequence} of cluster
ellipticals is well defined. The \emph{bright galaxies}, as defined above, 
are plotted as large symbols. These mainly follow the colour-magnitude
sequence for early-type galaxies within the cluster, which indicates that
their identification as members is likely to be correct.

\begin{figure}
  \centering
  \includegraphics[angle=-90,width=0.48\textwidth]{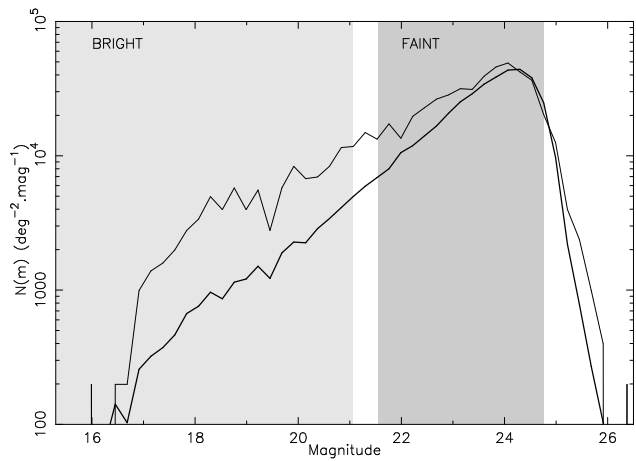}
  \caption{Number counts of galaxies in the Abell~1689 R-band image
    in bins of 0.23 magnitude. The thick line corresponds to counts of 
    galaxies across the whole field, the thin line to galaxies within
    $300\arcsec$ from the cluster centre. The greyed area to the left
    (right) shows the magnitude selection for
    bright (faint) galaxies.  }
  \label{fig:magdistrib}
\end{figure}

 A second catalogue is created for the faint galaxies, with the following
limits: $m^* + 2.5 < m < m^{\rm c} + 0.5$ for the B and I-band catalogue 
and $m^* + 3.5 < m < m^{\rm c} + 0.5$ for the R-band catalogue ($m^{\rm c}$
is the completeness magnitude which varies from filter to filter, see 
Table~\ref{tab:a1689params}). These catalogues are dominated by faint and 
hence probably distant galaxies and are therefore considered as catalogues 
of background galaxies lensed by the cluster. The different cuts were 
adjusted in order to separate the bright (foreground) and faint 
(background) galaxies as much as possible without losing too many galaxies
(see Fig.~\ref{fig:magdistrib}).

\section{Galaxy shape measurements}
\label{sec:psftxt}

 The shapes of stars detected in the images provide our best estimate of 
the point spread function (PSF), measuring the response of the entire
optical system (atmosphere + telescope optics) to a point-source. The
shape of a star includes an isotropic component mainly due to atmospheric
seeing, as well as an anisotropic component caused, for example, by
small irregularities in the telescope guiding. The isotropic component
of the PSF leads to a circularization of the images of small galaxies
and thus reduces the amplitude of the measured shear. The anisotropic
PSF component introduces a systematic component in galaxy
ellipticities and thus causes a spurious shear measurement if not
corrected \citep{kaiser95}. The geometric distortions of the camera
and the corresponding instrumental shear are corrected during the
data-reduction procedure when the image is reconstructed on a linear
tangential projection of the sky on a plane.

\begin{figure}
  \centering
  \includegraphics[angle=-90,width=0.48\textwidth]{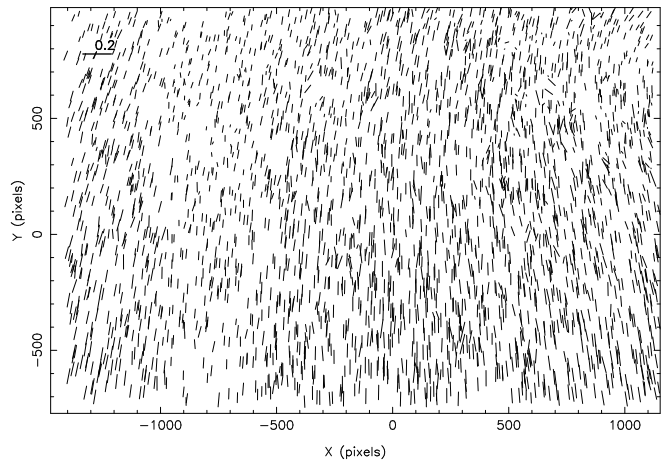}
  \caption{Abell~1689 R-band PSF map. The length of the vectors is
    proportional to their ellipticity as indicated by the scale in the
    upper-left corner. The origin of the figure is the cluster centre.
    See details in \S\ref{ssec:psfmaptxt}.}
  \label{fig:psfmap}
\end{figure}

 In the case of Abell~1689, which is representative of the entire survey,
the mean anisotropy of the PSF expressed in terms of ellipticity 
$\epsilon=(a-b)/(a+b)$ is much smaller than 0.15 in each filter (see
Fig.~\ref{fig:psfmap}).

In order to correct for both the PSF circularization and the PSF
anisotropy, we use the \textsc{im2shape} software developed by
\citet{bridle01}. \textsc{im2shape} implements a Bayesian approach to
measure the shape of astronomical objects by modelling them as the sum
of elliptical Gaussians, convolved by the local PSF, which is also
parameterized in terms of elliptical Gaussians. The minimization
procedure of \textsc{im2shape} estimates the posterior probability
distribution of the image given the model and the PSF, and Markov
Chain Monte Carlo sampling gives the most probable value for each
parameter, with the errors linked to the dispersion of the samples.
This approach is a practical implementation of the idea presented by
\citet{kuijken99}.  \textsc{im2shape} is becoming increasingly
popular, and has been used in a number of weak lensing applications using different
instruments \citep{Kneib2003, Cypriano2003, Faure2004}.

\begin{figure}
  \centering
  \includegraphics[angle=-90,width=0.48\textwidth]{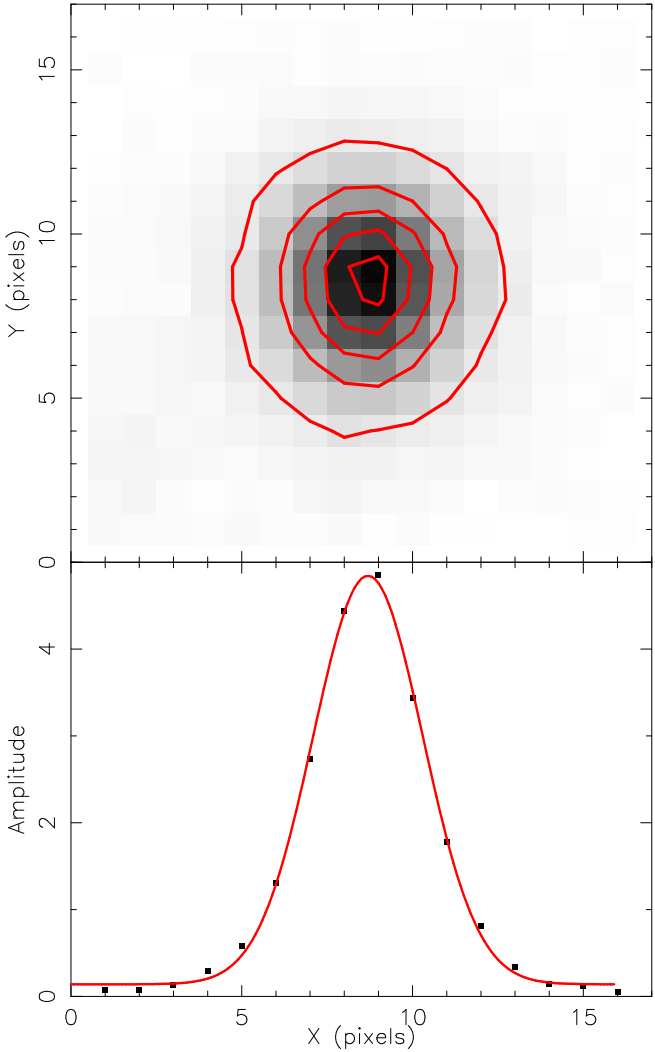}
  \caption{Top: a $16\times 16$ pixel
   image showing the averaged shapes of the five
    nearest stars to an arbitrary position (2000,2000)  
    in  the R-band image of Abell~1689. The contour levels are from 0.5
    to 4.5 in steps of 1. Bottom: A cut along the $x$-axis of the
    image above, indicated by small squares, and a Gaussian profile fit
    (obtained by \textsc{im2shape}) shown by the solid line.}
   \label{fig:psfprofile}
\end{figure}

 A detailed comparison between \textsc{im2shape} and the KSB method is
discussed by Bridle et al.\ (in prep.). In the following we describe in
detail the procedure we implement to transform the catalogue data into
source ellipticity parameters useful for a weak lensing inversion. For 
simplicity, only one elliptical Gaussian is used to describe both the shape
of the stars and the galaxies. Indeed, as shown in 
Fig.~\ref{fig:psfprofile}, star profiles are well fitted by a single
Gaussian. Furthermore, orientation and ellipticity (the most useful
parameters for the weak lensing analysis) are relatively insensitive to the
model used to describe luminosity profiles. The {\em a posteriori} 
justification of the validity of the choice is demonstrated by the quality
of the weak lensing measurements.

\subsection{Mapping the PSF distribution over the field}
\label{ssec:psfmaptxt}

In a first step, \textsc{im2shape} is used to measure the local PSF by
estimating the shapes of all the stars in the star catalogue. The
resulting PSF catalogue is then inspected in detail. We first remove
objects with ellipticity greater than 0.2 which mainly appear to be
defects between the chips. A second cleaning 
pass is done to remove stars
which are very different from their neighbours: 
if they are $>2\,\sigma$ away from
the mean value of the local seeing, they are automatically rejected
from the PSF catalogue.  The final cleaned distortion map measured
from the stars in the field is
presented in Fig.~\ref{fig:psfmap}.

\begin{figure}
  \centering
  \includegraphics[angle=-90,width=0.48\textwidth]{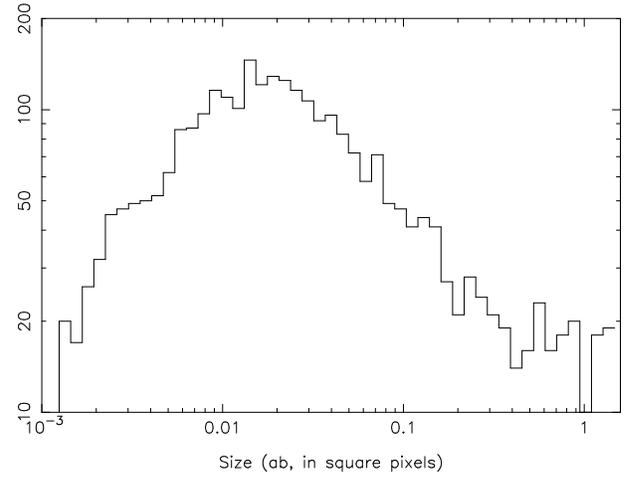}
  \caption{Left: The size distribution ($a \times b$) of the deconvolved
    stars. Their average size is consistent with them being point-sources.}
  \label{fig:stardeconv}
\end{figure}

\subsection{Faint galaxy shapes}
\label{ssec:faint_galaxy_shapes}

In a second step, we linearly interpolate 
the local PSF at each galaxy position 
by averaging the shapes of the five closest stars 
(Fig.~\ref{fig:psfprofile}). This number of stars is large enough to
locally interpolate the PSF, whereas a much larger number would over-smooth
the PSF characteristics.  The efficiency of the PSF measurement and
interpolation can be directly tested on the star catalogues. Fig.~\ref{fig:stardeconv} shows the resulting distribution of the intrinsic
sizes of stars after deconvolution with the local PSF. They are
intrinsically much smaller than 0.1\textsuperscript{th} of a pixel.

\textsc{im2shape} then computes the intrinsic shapes of galaxies by
convolving a galaxy model with the interpolated local PSF, and determine
which one is the most likely by minimizing residuals. In the
end, \textsc{im2shape}'s output gives a most likely model for the
fitted galaxy characterized by its position, size, ellipticity and
orientation, and errors on all of these.

\begin{figure}
  \centering
  \includegraphics[angle=-90,width=0.48\textwidth]{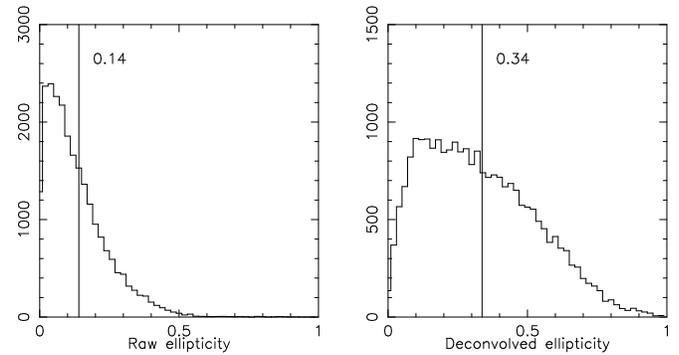}
  \caption{The ellipticity distribution of the faint galaxies in the
    R-band image of Abell~1689. Left: \textsc{im2shape } catalogue with
    no PSF correction. Right: \textsc{im2shape } catalogue with PSF
    correction.  The vertical line indicates the average value of the
    ellipticity.  The effect of circularization on the faint galaxies
    can easily be seen.}
  \label{fig:elldistrib}
\end{figure}

Fig.~\ref{fig:elldistrib} shows how the galaxy ellipticity
distribution alters after the \textsc{Im2shape} correction; the
effect of PSF circularization is evident.

\subsection{Mean redshift of the faint galaxies}
\label{ssec:meanzgal}

Although the photometric catalogues do not contain redshift
information on the background sources, we attempt to estimate this
for the population in a
statistical sense. This is necessary as the
relative distance of the background population and the lensing
cluster is of prime importance in
the quantitative scaling of the mass distribution from our  weak lensing
analysis. The critical parameter is the mean
value of $\beta=D_{\rm LS}/D_{\rm OS}$:
\begin{equation}
  \bar \beta = \frac{1}{N} \sum_{i=1}^{N} \frac{D_{{\rm LS},i}}{D_{{\rm
        OS},i}}
\end{equation}
where $N$ is the number of faint galaxies in the catalogue and $D_{\rm
  LS}$ is the angular diameter distance between the lens and the
source and $D_{\rm OS}$ between the observer and the source.

To compute $\bar \beta$ we have used a photometric redshift
catalogue produced from the {\it Hubble Deep Fields} (HDF) North and
South, observed with the {\it Hubble Space Telescope} (HST)
\citep{fernandez-soto99,vanzella01}. This catalogue, kindly provided
to us by S.\ Arnouts (priv.\ comm.), gives for each object in the
HDF-N/S the apparent magnitudes and colours
as well as measured spectroscopic redshift if it exists
\citep{vanzella02} or a photometric redshift otherwise. Similarly,
each galaxy in our three CFH12k catalogues (B, R or I) has at
least one entry in the corresponding photometric catalogues.
Depending on the number of available entries for each galaxy (1, 2
or 3) an automatic search is done in
the full HDF catalogues for the ten most similar objects in terms of
magnitude and colours (correcting for the slight differences between
the photometric systems of the CFH12k and WFPC2 cameras).  
Then the average redshift (photometric or
spectroscopic if available) of these 10 objects is assigned to the
galaxy. When photometric measurements
are available in all three filters
for an object, then this procedure crudely mimics 
a photometric redshift estimate, while it is a simple statistical average of
photometric redshifts at a given magnitude limit otherwise. Finally, the
mean redshift of each catalogue is computed, as well as the mean
$\bar{\beta}$. Their values are listed in Table~\ref{tab:a1689params}.

\section{Shear measurements}
\label{sec:shear_measurements}

 We have now measured the ``true'' shapes of faint galaxies and estimated
their mean redshift. The lensing equation for galaxy shapes can be written 
as:
\begin{equation}
  \label{eq:wl-nonlinear}
  \vec{\epsilon}_{\rm I} = \frac{\vec{\epsilon}_{\rm S} + \vec{g}}
  {1 + \vec{g}^* \vec{\epsilon}_{\rm S}}\quad,
\end{equation}
where $\vec{\epsilon}_{\rm I}$ and $\vec{\epsilon}_{\rm S}$ are the
complex ellipticities of the image and the source;
$\vec{g}=\vec{\gamma}/(1- \kappa)$ is the reduced shear;
$\vec{\gamma}$ is the shear vector and $\kappa$ is the convergence
\citep[e.g.\ ][]{Mellier1999,bartelmann01}. Note that both $\vec{\gamma}$
and $\kappa$ are proportional to the distance ratio $\beta$.
In the weak regime $\vec{g} \ll 1$ the above equation simplifies to:
\begin{equation}
  \label{eq:wl-linear}
  \vec{\epsilon}_{\rm I} = \vec{\epsilon}_{\rm S} + \vec{g}
\end{equation}

Assuming that the faint galaxy population lies at our estimated mean
redshift, and assuming that galaxies have random orientations in the
source plane, it is easy to see that by locally averaging a number
of ellipticities we have an unbiased estimate of the reduced shear,
allowing us to directly measure  the mass
distribution:
\begin{equation}
  \langle\vec{\epsilon}_{\rm I}\rangle = \langle\vec{g}\rangle
\end{equation}
The bracket $\langle\rangle$ indicate the average of a quantity
near a position. However, because of the random orientation of the galaxies
in the source plane, the error in the observed galaxy ellipticities and
thus on the estimated reduced shear will depend on the number of galaxies
averaged together to measure the shear \citep{schneider00}:
\begin{equation}
  \sigma_{\vec{g}} = \sigma_{\epsilon_{\rm I}}
                   \approx \frac{(1 - |g|^2) \, \sigma_{\epsilon_{\rm S}}}{\sqrt{N}}
  \label{eq:errshear}
\end{equation}
where $N$ is the number of galaxies used in the averaged.
$\sigma_{\epsilon_{\rm S}} \sim 0.33$ (see Fig.\ \ref{fig:elldistrib}) is
the dispersion of the intrinsic ellipticity distribution, and the
$(1 - |g|^2)$ factor is the effect of the shear on this dispersion. In the
weak lensing regime, $g$ is much smaller than 1 (our measurements reach 0.1
typically, see Fig.\ \ref{fig:rawprofile}), and this factor can be
neglected. The error on the shear measurement is then:
\begin{equation}
  \sigma_{\vec{g}} \approx \frac{\sigma_{\epsilon_{\rm S}}}{\sqrt{N}}
\end{equation}

Next we will explore different ways to do this averaging and constrain the
cluster mass distribution.

\subsection{Building the 2D shear map}
\label{ssec:shearmaptxt}

\begin{figure}
  \centering
  \includegraphics[angle=-90,width=.48\textwidth]{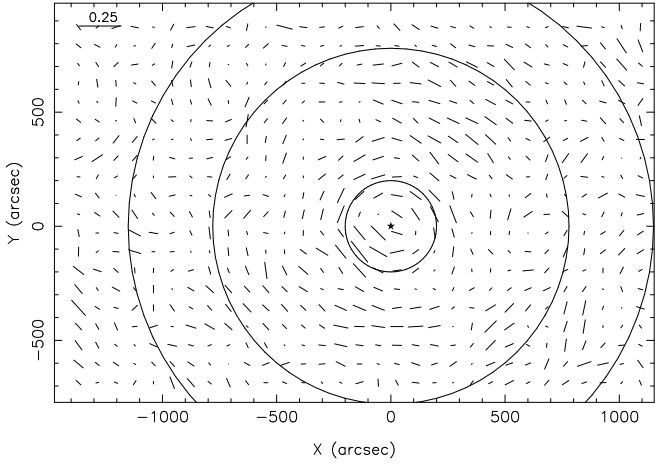}
  \caption{The shear map derived from the R-band image of Abell~1689. 
	  The cluster
           centre is marked by a ``$\star$''. The circles have radii of
           200\arcsec ($\sim 620\,\mathrm{kpc}$), 780\arcsec
           ($\sim 2.4\,\mathrm{Mpc}$) and 1200\arcsec
           ($\sim 3.7\,\mathrm{Mpc}$) respectively. The inner circle
           corresponds to the region of strong lensing, the second one to the
           largest circle that lies entirely within the CFH12k field, and
           the outer circle marks the limit where the area outside the
           field becomes significant.  The shear vectors are computed in
           cells of $80\arcsec \times 80\arcsec$, and have been smoothed
           by a Gaussian of 30\arcsec width (see details in
           \S\ref{ssec:shearmaptxt}).}
  \label{fig:shearmaps}
\end{figure}

The first and simplest test of the lensing influence of
Abell~1689 is to compute the 2D shear maps.  To
compute the \emph{shear maps} we average the galaxies in cells
using the lensing catalogue (PSF-corrected faint galaxy catalogue).
The cell size is chosen so that each cell contains about 35 galaxies.
At the magnitude depth of the catalogues ($\sim 20$ galaxies arcmin$^{-2}$) 
this number is typically achieved for cells sizes of $80\arcsec
\times 80\arcsec$.  Averaging such a number of galaxies
should mean that the measured average
ellipticity should be small (below 0.03 from Eq.\ 
\ref{eq:errshear}) and its orientation random in regions with no shear
signal. Near mass peaks, we expect to see the shear vectors
tangentially aligned around the centre of mass.  
Fig.~\ref{fig:shearmaps} clearly shows that 
we detect this characteristic lensing signal
around the cluster core in the R-band catalogue of
Abell~1689. The signal traced by the coherent alignment of the
``average'' galaxy shape is represented by vectors whose length is
proportional to the ellipticity and whose orientation follows the mean
orientation of the galaxies in each cell. Similar shear maps are seen
in the two other bands.

\subsection{Reconstructing the 2D mass map}
\label{ssec:lensenttxt}

We use the \textsc{lensent2} code \citep{marshall02} to compute the 2D
non-parametric mass map of the cluster. \textsc{lensent2} implements
an entropy-regularized maximum-likelihood technique to
derive the mass distribution within the field on a grid. 
The technique consists of a
Bayesian deconvolution process: a trial mass distribution
$\Sigma(\vec{\theta})$ is used to generate a predicted 
reduced shear field through the convolution of the surface mass
density by a kernel \citep[KS93:][]{Kaiser-Squires1993}. In
contrast to KS93, 
\textsc{lensent2} cannot produce negative feature 
in the mass maps leading to more physical solutions than can be 
obtained from direct reconstructions of the gravitational potential $\psi$.
Moreover, \textsc{lensent2} can include information not only from the
mean shear field, but from each individual lensed galaxy with its redshift (if 
known). As clusters of galaxies have smooth and extended mass 
distributions, the values of $\Sigma$ on the field are expected to be 
correlated through a kernel called the Intrinsic Correlation Function 
(\textit{ICF}). In our analysis, we provide
\textsc{lensent2} with a position, elliptical shape parameters (with
errors)  and an estimate of the redshift for each 
 lensed galaxy (we use the mean redshift as
explained in \S\ref{ssec:meanzgal}).  There is then only one
free parameter in the proceedure, the Intrinsic Correlation Function
(ICF) which measures the correlation between mass clumps.  We choose a
Gaussian ICF, and let its width vary. The ICF size is optimized so
that the reconstructed mass map does not contain a large number of
insignificant small-scale fluctuations, although small ICFs best fit
the mass peak of the cluster, while large ones best fit the wings of
the extended profiles. This optimization is performed by maximizing
the \emph{evidence} value of each reconstruction, which is the
probability to observe these data for a given ICF width. For more
details on \textsc{lensent2} see \citet{marshall02}.

\begin{figure}
  \centering
  \includegraphics[width=0.45\textwidth]{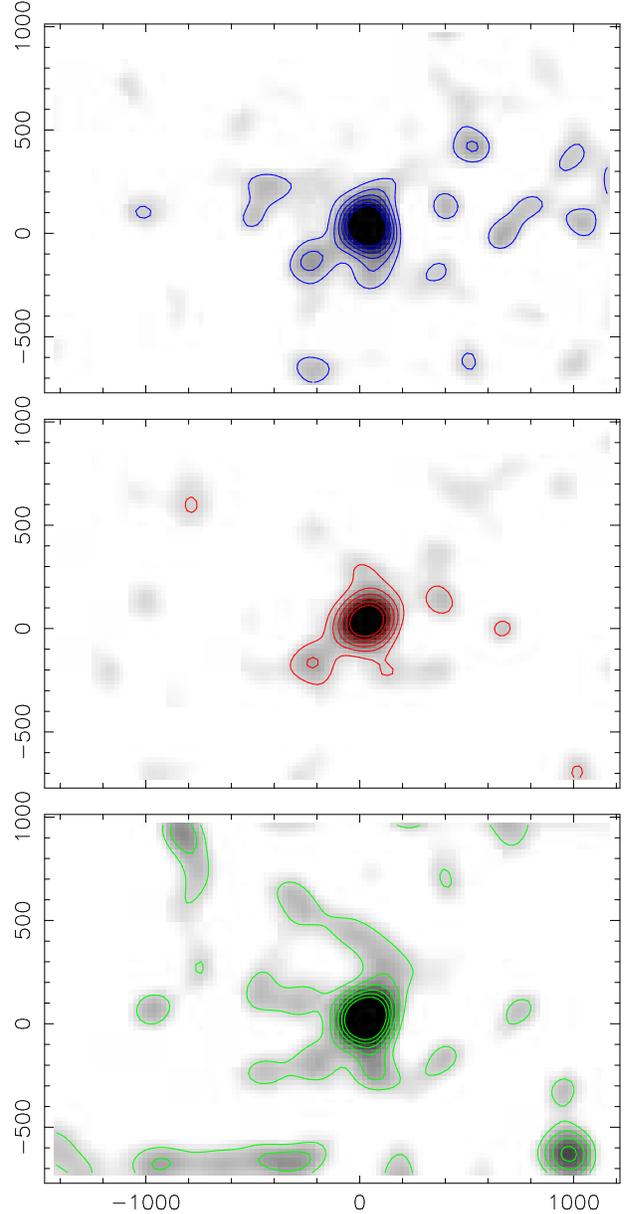}
  \caption{\textsc{LensEnt2} mass reconstructions for Abell 1689 from the
           B (top), R (middle) and I (bottom) catalogues. The ICF is
           Gaussian with a width of 180\arcsec. The cluster peaks are at
           1320, 1250 and $1090 \mpc2$ for B, R and I respectively. White
           (black) in the gray scale is set to $1\sigma$ ($5\sigma$), and
           contours are at 2, 3, 4 and $5\,\sigma$. $\sigma$ values are
           estimated as explained in \S\ref{ssec:lensenttxt}. The scale is
           in arcsec relative to the cluster centre. A possible secondary 
           peak is visible at [$-200\arcsec, -150\arcsec$].}
  \label{fig:panmassmaps}
\end{figure}

The main cluster mass clump is very well detected by
\textsc{lensent2}.  The code estimates the central surface mass
density of the peak, and gives its spatial configuration. Note that
large ICFs smooth the main peak.
Reconstructions are computed for a large set of ICFs (with
scales from $60\arcsec$
to $240\arcsec$), and the best ICF width is found to be near
160--$180\arcsec$ for our dataset.
An illustrative example is shown in Fig.~\ref{fig:panmassmaps} 
where the peak of the surface mass density is at
a value of $1250\mpc2$ in the adopted cosmology, although typical values
of the critical surface mass density for massive clusters at $z_{\rm L}
\sim 0.2$ are roughly around $\Sigma_{\rm c} = 3200\mpc2$ for sources at
$z_{\rm S} \sim 1.0$. This is because the ICF width used here
($180\arcsec$) is much larger than the  Einstein
radius of the cluster ($\sim\,40\arcsec$). Therefore the smoothing process
strongly attenuates the central peak density which in the case of
Abell~1689 is clearly over-critical.

 To assess the significance of the other mass density peaks detected in
each image we randomize the orientation of the faint galaxies in the 
lensing catalogue, while keeping their positions and axial ratios fixed. We
perform mass reconstructions of 200 randomized catalogues, and in each
identify the 15 highest significance mass peaks. The statistics of these
3000 values gives a mean noise peak of $116 \mpc2$ (99, 85) above the 
background level (set at $100 \mpc2$ in input of \textsc{lensent2}) 
respectively in the R (B, I) images. This value is considered as the 
average fluctuation of the noise peaks, $\sigma$. With this definition, 
the cluster mass peak is detected at nearly $10\sigma$ above the 
background. To be formally correct, prior to randomizing their orientations
we should also ``unlens'' the galaxies using the shear determined above and
applying Eq.\ref{eq:wl-nonlinear}. This has not been done yet for 
simplicity and will be explored in more detail in the next paper (Bardeau 
et al.\ in prep, paper II). However, as the lensing induced distortion is
almost always very small compared to the width of the ellipticity 
distribution, we do not expect that this simplification will affect the
estimated significance of the mass peak.

 \textsc{lensent2} mass reconstructions give many low significance mass
peaks. For example, Fig.~\ref{fig:panmassmaps} shows that four clumps reach
the 2-$\sigma$ level, but only one is above the 3-$\sigma$ level (excluding
the main cluster peak). To check their reality, we can compare the 
reconstructions derived from the three filters (B, R and I, 
Fig.~\ref{fig:panmassmaps}).  The regions where a mass clump is detected in
all three images are considered as ``real'' ones and can be compared to the
number density map of bright cluster galaxies. Another test is to compare 
these clumps with any enhancement of the light distribution 
(\S\ref{ssec:lightmap}), provided that the mass clumps are not associated 
with ``dark clumps''. The multi-colour approach in our weak lensing survey
provides a powerful tool to eliminate most of the inconsistencies created 
in the mass reconstructions from defects in the lensing catalogues. For
Abell~1689, apart from the mass peak associated with the cluster, no other 
$>3\sigma$ peaks were detected in all three filters. A possible 
2--3$\sigma$ peak is located 5\arcmin South-East of the cluster but no
obvious counterpart in the galaxy distribution can be identified.

\subsection{The radial shear profile}
\label{ssec:profiles}

\begin{figure*}
  \centering
  \includegraphics[angle=-90,width=0.85\textwidth]{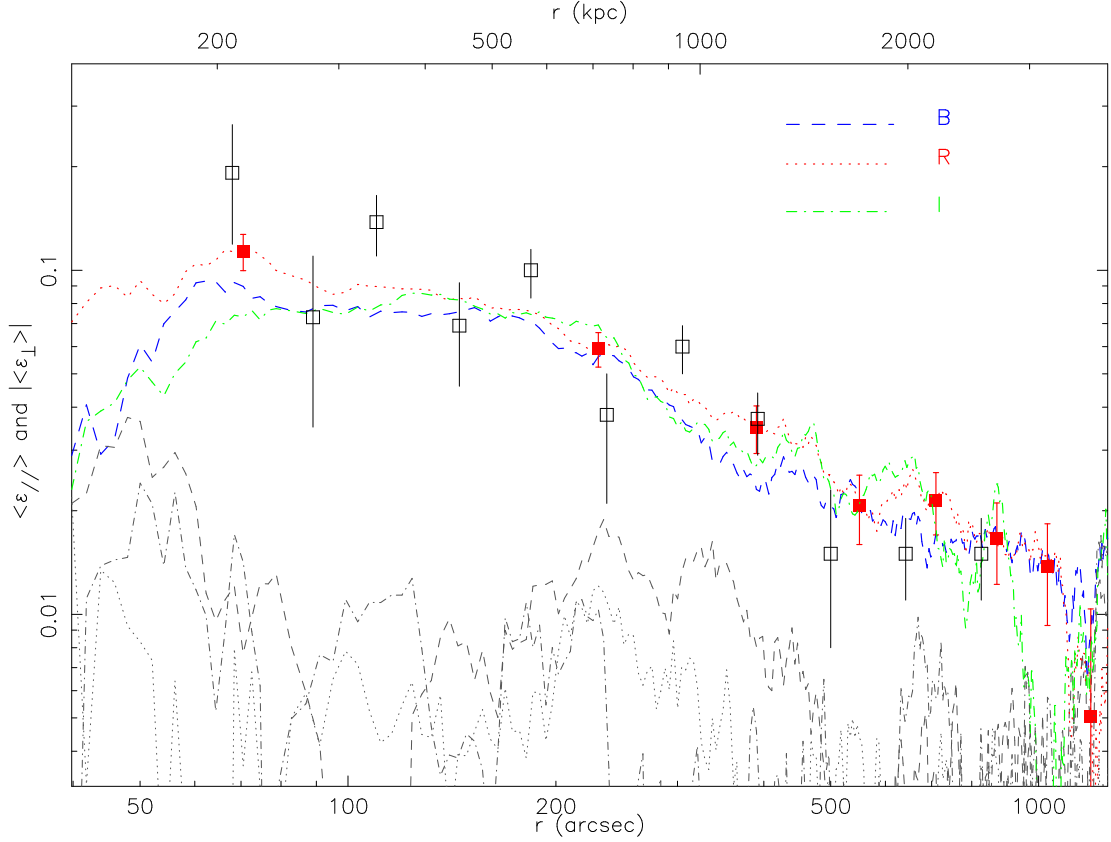}
  \caption{The tangential shear profile for Abell~1689, in the B, R and I bands. 
    The bin width ($\Delta R$) is $160\arcsec$. A series of uncorrelated points with error bars is
    displayed for the R band (solid squares). Absolute values of the
    associated radial component $|<\varepsilon_\perp>|$
    are indicated in grey at the bottom of the plot, showing that
    the signal is well detected out to $\sim 1000\arcsec$ ($\sim3$ Mpc) from
    the centre. The measurements of \citet{clowe01} are also shown for
    comparison (open squares).}
  \label{fig:rawprofile}
\end{figure*}

We have demonstrated that only one significant mass peak is detected
in the field of Abell~1689, and that it aligns well with the
cluster centre indicating it corresponds to the potential
well of the cluster. In order to quantify the mass of this clump we
analyse the radial distribution of the shear around this peak. Tangential
and radial shear components are computed as a function of the distance to the
cluster centre.  They are averaged in annuli of width $\Delta R = R_2
- R_1$ for a mean radius $R= \left( {R_1} + {R_2} \right) / 2$.
$\Delta R$ is kept constant so the S/N of the shear roughly decreases
as $1/\sqrt{R}$, in order to keep enough independent points at large
radii (a constant S/N requires too large annuli at these radii).  A
quasi-continuous profile is built by using a ``sliding window'' with
steps $\Delta r$ much smaller than $\Delta R$.  In practice, we chose
$\Delta R = 160\arcsec$ (and $\Delta r=10\arcsec$) for the Abell~1689
R image, so about 10 independent points are included in the profile.

Fig.~\ref{fig:rawprofile} shows the tangential and radial shear
profiles from the three images of the field. 
The radial shear should be zero in the
case of perfect data and a well chosen centre for the annuli.  In
practice, it can be considered as an independent estimator of
measurement errors (this is also referred to as the 45 degree test).
In the case of Abell~1689, the radial shear is always lower than the
tangential shear out to $\sim 1100\arcsec$, arguing for good data
quality in all three bands.

Note that in the very centre ($R < 70\arcsec$) the shear profile appears
to drop.  The error bars are large due to the low number statistics:
the area considered is small, reduced still further by the masking effect of the
bright galaxies. Moreover the depletion of the number density of
background galaxies in the center due to the magnification bias
\citep{Taylor1998} also decreases the number of observable galaxies,
although this effect is only important in the inner-most annuli. These
low number statistics does not completely explain the weakness of the
shear: it can also be under-estimated if unlensed galaxies (such as
cluster members) are included in the catalogues, which should be more
likely towards the cluster core.  As a consequence, the points inside
$R=70\arcsec$ will not be used in the modeling of the shear profile.
The measurements done by \citet{clowe01} using R band images from the ESO
Wide Field Imager (WFI) are also presented in Fig.~\ref{fig:rawprofile}
for comparison. Our measurements are quantitatively in good agreement with
those of \citet{clowe01}. Moreover, our error bars are smaller and our
points less scattered, even if we consider the different binnings. This
strongly suggests that the use of \textsc{im2shape} in the analysis
process improves significantly the shear measurements. This will be
quantified in a forthcoming paper (Bridle et al.\ 2005, in preparation).

\section{Modeling the lensing data}
\label{sec:modelling}

\subsection{Description of the mass models}
\label{ssec:mass_models}

Three families of mass models are used to fit the measured shear
profile: a singular isothermal sphere profile (SIS), a power law
profile (Pow) and finally the ``universal'' NFW profile
\citep{Navarro1997}. In addition we implemented the Aperture Mass
Densitometry method (AMD) to compute a \emph{non-parametric} mass
profile from the shear profile itself \citep{fahlman94}.
We recall briefly the basic equations for the mass density ($\rho$),
shear ($\gamma$) and convergence ($\kappa$) profiles for the three
models.

\subsubsection{The Singular Isothermal Sphere model}
\label{ssec:SIS}

This is the simplest mass profile used in lensing inversion. It is
essentially given by the following equations:
\begin{eqnarray}
  \label{eq:SISdef}
  \rho(r)        & = & \frac{\sigma^2}{2 \pi G r^2} \\
  \kappa(\theta) & = & \gamma(\theta) = \frac{\theta_{\rm E}}{2 \theta} \\
  \theta_{\rm E} & = & \frac{4 \pi \sigma^2}{c^2}
  \frac{D_{\rm LS}}{D_{\rm S}}\quad,
\end{eqnarray}
where $\sigma$ is the velocity dispersion of the
cluster.  Note that once the cluster centre is fixed, this profile
depends has only one free parameter ($\theta_E$ or equivalently $\sigma$),
so only one degree of freedom is available in the fits.

\subsubsection{The Power Law model}
\label{ssec:Pow}

The Power Law model is a generalization of the SIS model, where the
slope of the mass density profile is a free parameter
\citep{schneider00}.
\begin{eqnarray}
  \label{eq:POWdef}
  \gamma(\theta) & = & \frac{q}{2}   \left( \frac{\theta}{\theta_{\rm
        E}} \right)^{-q}\\ 
  \kappa(\theta) & = & \frac{2-q}{2} \left( \frac{\theta}{\theta_{\rm
        E}} \right)^{-q}  
\end{eqnarray}
where $q$ is the slope of the Power Law ($q = 1$ for the SIS model).
Once the cluster centre is fixed, this model provides two degrees of
freedom for fitting.

\subsubsection{The NFW profile}
\label{ssec:NFW}

The NFW profile is derived from fitting the density profile of numerical
simulations of cold dark matter halos \citep{Navarro1995, Navarro1997}.
This theoretically-motivated profile is 
becoming increasingly popular in 
weak lensing analyses  of clusters \citep{Kneib2003} 
as it appears to give a reasonable description of the observed
shear profiles. The
mass density profile can be expressed as
\begin{eqnarray}
  \rho(r) & = & \frac{\delta_{\rm c} \rho_{\rm c}}{(r/r_{\rm
      s})(1+r/r_{\rm s})^2} \\
  \mbox{where} \,\,\, \delta_{\rm c} & = & \frac{200}{3} \frac{c^3}{\ln(1+c)-c/(1+c)} \\
  \mbox{and} \,\,\,\rho_{\rm c} & = &
  \frac{3 H^2(z)}{8 \pi G}\,.
\end{eqnarray}
$r_{\rm s}$ is the scale radius, $H(z)$ the Hubble parameter and $c =
r_{200}/r_{\rm s}$ the concentration parameter which relates the scale
radius to the virial radius $r_{200}$. This density profile is
shallower than the SIS near the center but steeper in the outer parts.
Similarly as the power law model, once the centre is fixed, it has two
degrees of freedom: $M_{200}$ for the normalization of the mass and
$r_{\rm s}$ for the scale radius, or equivalently $r_{200}$ and $c$.
The details of the analytic expressions for the shear and convergence
of the NFW profile can be found in \citet{King2002}.

\begin{table*}
  \centering
  \caption{Best fit results for the Abell~1689 R-band shear profile. For
    the SIS, the results are given in terms of Einstein radius
    ($\theta_{\rm E}$) and velocity dispersion $\sigma_{\rm los}$.
    For the Power Law, $\theta_{\rm E}$ is again the Einstein radius and
    $q$ the logarithmic slope. Finally for the universal NFW profile, $c$
    is the concentration parameter and $r_{200}$ the virial radius.
    $M_{200}$ is the 2D-projected mass inside $r_{200}$ in units of
    $10^{12}\,h_{70}^{-1}\,M_{\sun}$ and $\theta_{\rm E}$ is the
    derived Einstein Radius.
    (a) refers to the fit results from \citet{clowe01}, (b) from
    \citet{King2002}. The numbers in italics assume $z_{\rm s}=1.06$.
  }
  \begin{tabular}{cccccc}
\hline
\hline
\noalign{\smallskip}
SIS & $\sigma_{1D}\,(\mathrm{km\,s^{-1})}$ & $\theta_E (\arcsec)$ & & & $\chi^2$ \\
    & $ 998 \pm 68$                        & $22.4 \pm 3.0$       & & & 1.98 (1)\\
(a) & $1028 \pm 35$                        & ${\it 23.8 \pm 1.6}$ & & & \\
\noalign{\smallskip}
\hline
\noalign{\smallskip}
Pow & q               & $\theta_E (\arcsec)$ & & & $\chi^2$ \\
    & $0.75 \pm 0.07$ & $14.6 \pm 0.3$       & & & 0.637 (2)\\
(b) & 0.88            & 18.0                 & & &          \\
\noalign{\smallskip}
\hline
\noalign{\smallskip}
NFW &  c                   & $r_{200} \: ({h_{70}}^{-1}$ Mpc) & $M_{200}\:(10^{12}\:{\rm M}_{\sun})$ & $\theta_E (\arcsec)$ & $\chi^2$ \\
    &  $3.5^{+0.5}_{-0.3}$ & $1.99 \pm 0.25$                  & $1410^{+630}_{-470}$                 & $2.6^{+1.4}_{-0.2}$  & 0.334 (2)\\
(a) &  6.0                 & 1.83                             & {\it 1030}                           & {\it 9.7}            &          \\
(b) &  4.8                 & 1.84                             & {\it 1070}                           & {\it 5.3}            &          \\
\noalign{\smallskip}
\hline
\end{tabular}

  \label{tab:fittedvalues}
\end{table*}

\subsection{Weak lensing fit}
\label{ssec:WLfit}

 Each of the three models presented above is fitted to the data with a
least square minimization over the parameter space of the models.
The $\chi^2$ value is then:
\begin{equation}
  \chi^2 = \frac{1}{{\cal N}-1} \sum_{k=1}^{{\cal N}} \left(
    \frac{\epsilon^{\rm t}_k - g_{\rm model}(x_k)}{\sigma_k}
  \right)^2 \quad ,
\end{equation}
where $\cal N$ is the number of data bins, and $\sigma_k$ is the error on
the tangential ellipticity. The error is computed in each bin as the mean
error on the tangential ellipticity ($\epsilon^{\rm t}$), weighted by the
number $N_k$ of galaxies in the bin used to do the measurement:
$\sigma_k = \langle \sigma_{\epsilon^{\rm t}} \rangle_k / \sqrt{N_k}$.

 The data in the outer regions at $r>r_{\rm max}$, where the annuli reach
the borders of the field, are excluded. In practice, only the area where
the tangential shear is greater than radial shear is included in the fits.
Furthermore, as explained in \S\ref{ssec:profiles}, we also exclude the
central part of the cluster. In the case of the R-band observations of
Abell~1689, the fitted range corresponds to radii from
$r_{\rm min}=70\arcsec$ to $r_{\rm max}=1100\arcsec$.

\begin{figure*}
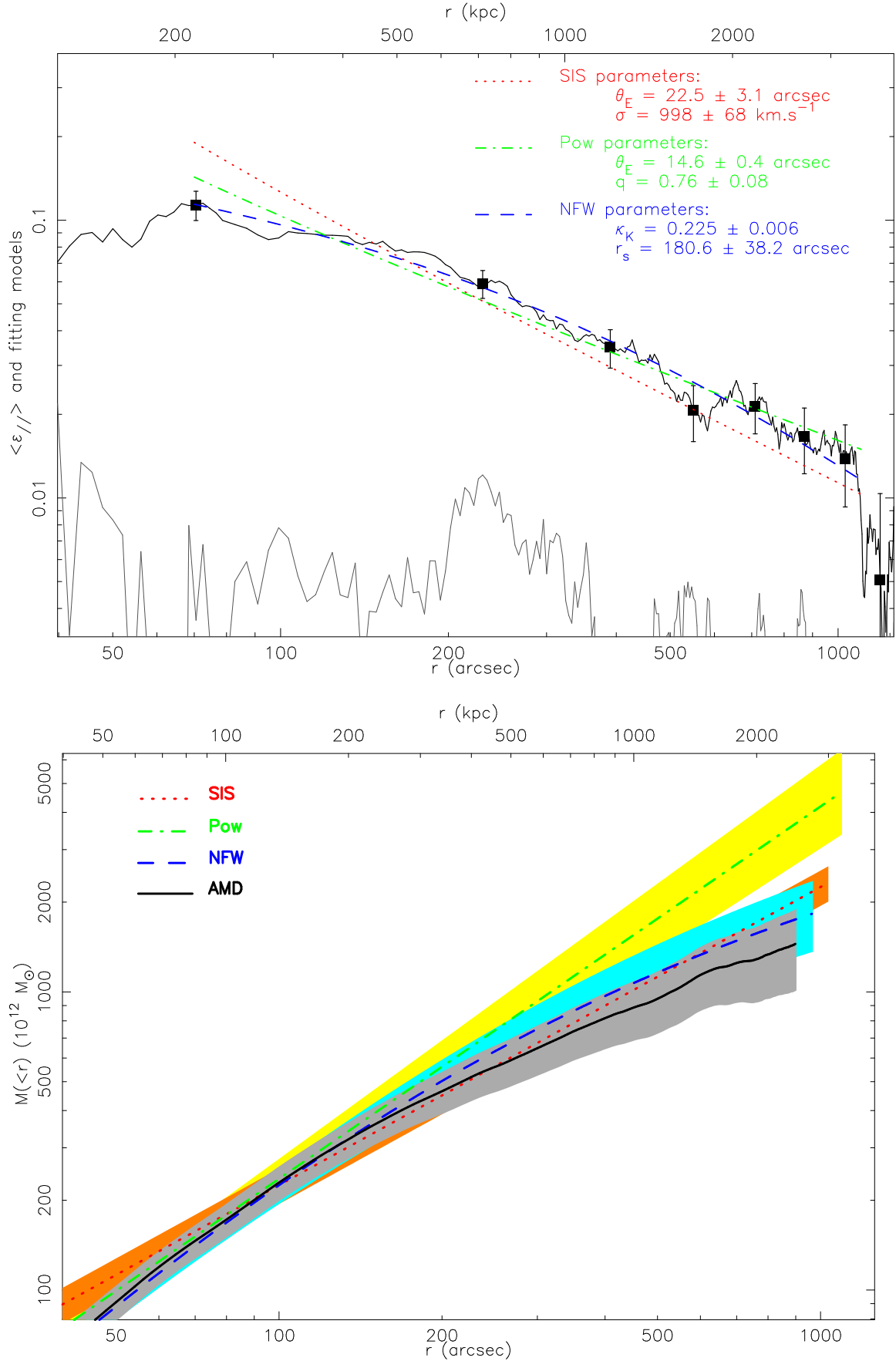

  \centering
  \includegraphics[angle=-90,width=0.85\textwidth]{1643f12a.ps}
  \includegraphics[angle=-90,width=0.85\textwidth]{1643f12b.ps}
  \caption{Top: Best fitting parameters for SIS, Power law and NFW
    models, for the Abell~1689 R-band shear profile. One series of
    uncorrelated points is shown (bin width $= 160\arcsec$).  Bottom:
    Deduced mass profiles from these models. The mass profile from the
    Aperture Mass Densitometry Method is also displayed, with a
    reference radius of 1100\arcsec. See text for details.}
  \label{fig:models}
\end{figure*}

Table~\ref{tab:fittedvalues} summarizes the results of the fits, and
Fig.~\ref{fig:models}a displays the resulting best-fit models. The lower
quality of the fit by the SIS profile is easy to understand as it depends
on one parameter only, contrary to the other two which are represented by two
parameters. Moreover, the value of the Einstein radius deduced from the
fit is significantly lower than that measured from strong lensing
(which is estimated to be $\theta_{\rm E}=41\arcsec$). Note that
\citet{clowe01} deduced from their weak lensing analysis a value for the
Einstein radius similar to our estimate.

The fit with a power law is slightly better than the
SIS as the slope of the
profile appears to be shallower than isothermal, but the Einstein
radius is again some 25\% lower than expected
(\citet{King2002} found similar
results with an even lower Einstein radius).

The universal NFW profile provides the best fit to our shear
profile.  The concentration parameter ($c$) is slightly smaller than
the values found by \citet{clowe01} and \citet{King2002},
whereas the virial radius $r_{200}$ is very similar. The derived
Einstein radius is however quite small and thus this model is not a
good fit of the central parts of the cluster.

We conclude that 
total mass profile of this cluster 
across all angular scales is not well described
by any of these simple fitting formulae and requires a more complex
model, perhaps including
contributions from the cluster galaxy halos and possibly a
steeper central mass distribution.

Fig.~\ref{fig:models}b shows the projected mass profiles from the
previous fits computed with the following equation
\begin{equation}
  M(r) = \pi r^2 \, \Sigma_{\rm c} \, \bar{\kappa}(r)
\end{equation}
where $\bar{\kappa}(r)$ is the mean dimensionless surface mass
density inside radius $r$.

\subsection{The Aperture Mass Densitometry method}
\label{ssec:AMD}

 Instead of fitting analytical formulae, we can directly integrate the
measured reduced shear to determine the relative mass profile within the 
cluster. This direct method has been developed by \citet{fahlman94} and is
called ``Aperture Mass Densitometry'' (AMD). The function $\zeta(r_1, r_2)$
is defined as the difference between the average convergences (or mean
projected mass densities) inside the radius $r_1$ and within the annulus 
between $r_1$ and $r_2$:
\begin{eqnarray}
  \zeta(r_1,r_2) & = & \bar{\kappa}(r < r_1) - \bar{\kappa}(r_1 < r < r_2) \label{eq:zeta}\\
                 & = & \frac{2}{1-\left( r_1/r_2 \right)^2} \:
                 \int_{r_1}^{r_2} \frac{\gamma_{\rm t}}{1-\kappa(r)}
                 \: d \ln r \quad .
\end{eqnarray}
 The reconstructed mass inside the radius $r<r_{\rm max}$ is then
\begin{equation}
  \label{eq:AMD-mass}
  M_{\zeta}(r) = \pi r^2 \,\Sigma_{\rm c} \; \zeta(r,r_{\rm max}) \quad ,
\end{equation}
where $r_{\rm max}$ is the maximum radius for which we can measure the
shear or the radial limit of the data. In the case of our observations
of Abell~1689, we choose $r_{\rm max} \sim 1100\arcsec$, the maximum
radius where annuli lie entirely within the field of view.
Regarding Eq.\ \ref{eq:zeta}, $M_{\zeta}(r)$ is only a lower limit to
the \emph{true} mass $M(r) = \pi r^2 \,\Sigma_{\rm c} \,
\bar{\kappa}(r)$ and should not be considered as an absolute mass
determination.

 The AMD mass profile is shown in Fig.~\ref{fig:models} with the mass
profiles derived by fitting the various analytical models. As expected, we
find that the mass estimated from AMD is always lower than the parametric 
mass estimates.

\section{Light distribution and mass-to-light ratio}
\label{sec:light-distribution}

\subsection{2D light distribution}
\label{ssec:lightmap}

 The catalogue of ``bright'' galaxies is expected to be dominated by
cluster members, although it may also contain other bright galaxies within
the field of view. Thus a density map (light density or number density) 
derived from this catalogue can trace the morphology of the cluster and any
associated structures in its vicinity. In the case of Abell~1689, no galaxy
over-densities, other than the main cluster component, are associated with 
any prominent peaks in the lensing mass distribution 
(Fig.~\ref{fig:image}).

 We therefore focus on the distribution of light around the cluster centre
assuming that the observed over-density is due to cluster members. First in
order to build a quantitative light density map or its radial profile, it
is necessary to statistically correct the catalogue for the field 
contamination. Fortunately, the CFH12k images are large enough so that at 
radii beyond 600\arcsec\ from the cluster centre (2 Mpc at the cluster 
redshift) we can assume that the galaxy density is close to the ``field'' 
density. The mean number and light densities are therefore corrected by 
subtracting their minimal values estimated in the area 
$600\arcsec < R < 1200\arcsec$.

 Furthermore in order to estimate the total luminosity of the cluster and
its radial profile, it is necessary to correct for the magnitude limit of
the catalogue, corresponding to a cut in the cluster luminosity function
(LF). The incompleteness factor $C$ is estimated as follows, the cluster LF
is assumed to follow the standard Schechter luminosity function 
\citep{schechter76}:
\begin{equation}
  \label{eq:Schechter-LF}
  \phi(L) = \frac{{\rm d}N}{{\rm d}L} = \frac{\phi^*}{L^*} \
  \left(\frac{L}{L^*}\right)^{-\alpha} \ {\rm e}^{-L/L^*}
\end{equation}
Therefore the luminosity integrated in the catalogue down to a luminosity
$L_{\rm inf}$ is
\begin{eqnarray}
  \label{eq:Schechter-LF-integrated}
  L_{\rm cat} & = & \int_{L_{\rm inf}}^{+\infty} L \, \phi(L) \, {\rm d}L \\
              & = & \phi^* L^* \, \left[\Gamma(2-\alpha) - \Gamma(2-\alpha,L_{\rm inf}/L^*)\right]
\end{eqnarray}
so the fraction of the luminosity not taken into account when integrating
within the magnitude limits of the catalogues is written as
\begin{equation}
  \label{eq:luminosity_incompleteness}
  C = \frac{\Gamma(2-\alpha,L_{\rm inf}/L^*)}{\Gamma(2-\alpha)}
\end{equation}
and the total luminosity is
$\displaystyle L_{\rm tot} = L_{\rm cat} / (1-C)$.

 For the three bands used in this study, we need to estimate the two main
parameters of the Schechter luminosity function $\alpha$ and $L^*$. These
parameters depend on the choice of filters, on the mix of galaxy types, and
on the cosmological model.  The best multi-colour luminosity function
determinations are presently those from the Sloan Digital Sky Survey (SDSS)
early release data \citep{blanton01}, although they correspond to a field
LF. The SDSS photometric system ($u, g, r, i, z$) is transformed to the 
CFH12k (Johnson) system by applying the transformations of 
\citet{fukugita96}. In this paper we use the parameters of the LF 
summarized in \citet{delapparent03} and applied to a Sbc galaxy. Therefore
the absolute magnitude $M^*$ in the R filter is $-21.83$ in the adopted 
cosmology and the slope is $\alpha_{\rm R}=1.20$. This includes also the 
k-correction at redshift $0.18$, computed with the galaxy evolutionary code
by \citet{bruzual03}.

 Finally, the correction factors $1/(1-C)$ are applied to $L_{\rm cat}$ to
obtain the total integrated magnitude for the B, R and I catalogues, with
the magnitude ranges defined in Sect.~\ref{ssec:galscats}. The precise
values derived are summarized in Table~\ref{tab:lumfunct}.

\begin{table}
  \centering
  \caption{The photometric parameters of the luminosity function in B, R
           and I filters for our adopted cosmology: $H_0 = 70 \,
           \mathrm{km\,s^{-1}\,Mpc^{-1}}$, $\Omega_{\rm m} = 0.3$ and
           $\Omega_\Lambda = 0.7$. The distance modulus is $m-M=39.70$ or
           equivalently the luminosity distance is $D_{\rm L} = 872 \,
           \mathrm{Mpc}$, at $z_{\rm A1689}=0.18$. $1/(1-C)$ is the
           correction factor applied to the integrated luminosity of the
           catalogues to determine the total luminosity of the cluster.}
  \begin{tabular}{cccc}
    \hline
    \hline
    \noalign{\smallskip}
    & B & R & I \\
    \noalign{\smallskip}
    \hline
    \noalign{\smallskip}
    k-correction          & 1.06   & 0.16   & 0.16 \\
    $M^* - 5 \log h_{70}$ & -20.47 & -21.83 & -21.54 \\
    $\alpha$              & 1.30   & 1.20   & 1.25 \\
    $m^* \ (z=0.18)$      & 20.29  & 18.03  & 17.32 \\
    $1/(1-C)$             & 1.28   & 1.11    & 1.27 \\
    \noalign{\smallskip}\hline
  \end{tabular}
  \label{tab:lumfunct}
\end{table}

\begin{figure*}
  \centering
  \includegraphics[angle=-90,width=0.9\textwidth]{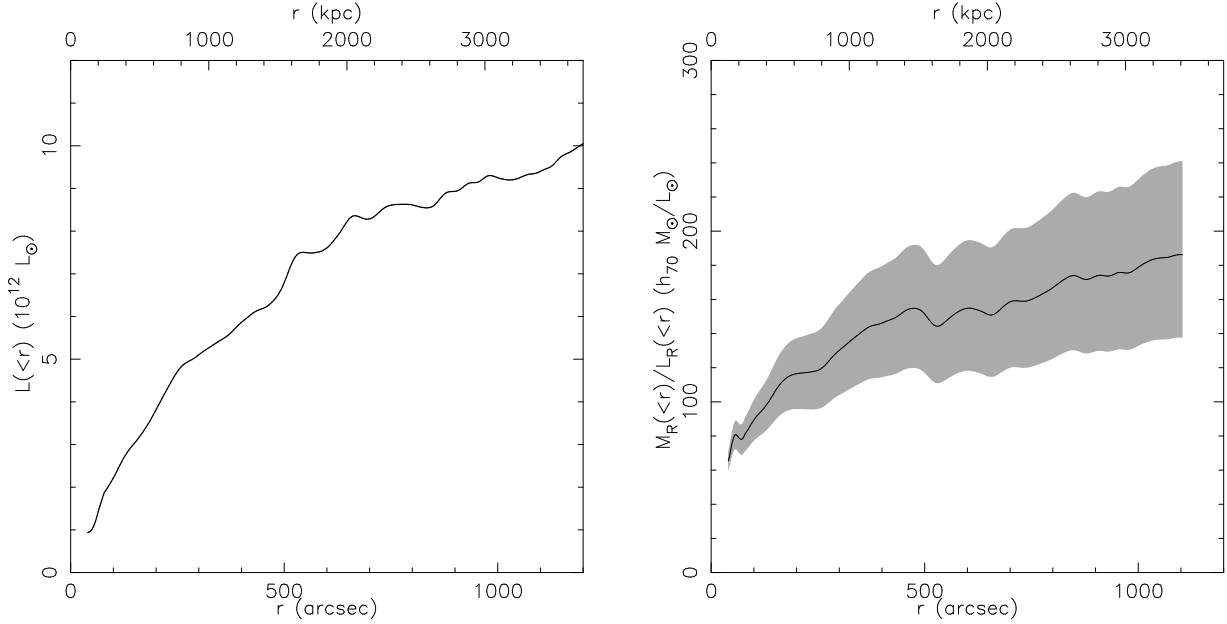}
  \caption{Left: luminosity profile for the bright galaxy catalogue, for
           R-band image of Abell~1689, corrected for background
           contamination. See Sect.~\ref{ssec:lightmap} for details. Right:
           $M_R/L_R$ ratio as a function of radial distance from the 
           cluster centre. The mass profile is estimated from the best fit
           NFW parameters. The filled region indicates the errors on the
           profile.}
  \label{fig:luminprof}
\end{figure*}

\subsection{Comparison of mass and light: M/L radial profile}
\label{ssec:mlprof}

 Using the best-fitting NFW model for  the observed shear profile, the 
$M/L$ profile is computed by dividing the luminosity profile, estimated 
from the bright galaxies catalogue, by the mass profile. The field
contamination in this catalogue is estimated by measuring the minimum of
the surface brightness density between $600\arcsec$ and $1200\arcsec$ from
the cluster center.  Fig.~\ref{fig:luminprof}a displays this integrated
luminosity profile for the R-band image. Note that the $L_{\rm R}$ values
adopt the correction factor discussed in the previous section.

 Fig.~\ref{fig:luminprof}b displays the $M_{\rm R}(<r) / L_{\rm R}(<r)$
profile with error bars estimated from the errors on the mass profile only. 
The $M/L$ increases from a low value (near $100\pm10 \, h_{70} \, 
(M/L)_{\sun}$ at $400\,\mathrm{kpc}$ from the centre) and flattens 
out beyond $\sim 1 \, {\rm Mpc}$ at a value near $160 \pm 40 \, h_{70} \,
(M/L)_{\sun}$. This behaviour is independent of the filter considered. 
It does however depend slightly on the background correction at large 
radius, and on the detailed mass modeling in the inner part of the cluster.
In particular, as we derived a relatively small Einstein radius compared to
that determined from strong lensing, we are likely to be underestimating
the mass in the central regions, which would suggest an even flatter $M/L$
profile towards the centre.

 Beyond $\sim1\,\mathrm{Mpc}$ the $M/L$ ratio found in Abell~1689 is
consistent with being constant with radius. This result is similar to the
findings of \citet{Kneib2003} in their lensing analysis of the cluster 
Cl\,0024+1654 ($z=0.39$), both in the radial distribution and in the
normalization. For comparison, $(M/L)_{\rm R}$ at large radii in the Coma
cluster is found to be $170\pm50\,(M/L)_{\sun}$ from dynamical 
analysis \citep{Geller1999, Rines2001}. Similar profiles for mass and light
on 1--5 Mpc scales are expected if cluster assembly is largely governed by
infalling groups and if no strong mass segregation occurs in the cluster.

 In their sample of 12 distant clusters ($0.17 < z <0.56$) \citet{smail97}
found a mean value of $(M/L)_V^{\rm all} = 126_{-77}^{+147} \, 
(M/L)_{\sun}$ ($h=0.7$) in the cluster cores, where the superscript
\emph{all} refers to the entire population of the clusters, not only 
early-type galaxies. Given the colour index ($V-R$) of a mean Sa galaxy at 
redshift 0.18, this corresponds to $(M/L)_R^{\rm all} = 102_{-62}^{+119}\,
(M/L)_{\sun}$. Since our bright galaxies catalogue is dominated by 
elliptical galaxies (Fig.\ref{fig:R-I_I}), we expect to find a lower 
luminosity thus their $M/L$ value is consistent with our findings.

\section{Discussion and Conclusion}
\label{sec:conclusions}

 In this paper, we describe the methodology used to analyze a multi-colour
wide-field imaging survey of 11 X-ray luminous clusters. The goal of our 
survey is to constrain the mass distribution in clusters of galaxies using 
weak gravitational lensing. The main elements of the data analysis are: the
use \textsc{SExtractor} for object detection and photometry to provide 
well-defined object catalogues. A ``stars'' catalogue is used to determine 
the PSF locally, a ``bright galaxies'' catalogue is defined to trace the 
distribution of cluster members and a ``faint galaxies'' catalogue is
constructed which should comprise background galaxies. The magnitude limits
of each catalogue are determined with respect to the observational
constraints such as the limiting magnitudes of the available images as well
as physical constraints related to the magnitude distribution in the
clusters at a given redshift. In order to determine the ``true'' 
PSF-deconvolved shape properties of the background (lensed) galaxies we use
the \textsc{im2shape} package developed recently for the purpose of 
improving the quality of shear measurements, including a correct treatment 
of the measurement errors \citep{bridle01}. We then reconstruct the mass 
distribution by computing the shear profile and either fitting it with 
parametric mass models like the NFW mass profile or deducing the relative 
profile directly with the non-parametric Aperture Mass Densitometry method.
Both methods are found to be consistent. We also show a 2D mass 
reconstruction using the {\sc lensent2} software \citep{marshall02} and 
applying it to the three images taken through the three filters. Finally we
compute the $M/L$ ratio as a function of radius, again in the three 
photometric bands. The three filters are used independently for most of the
processing steps in order to confirm the significance of the results 
(comparison of shear profiles and mass maps). They give quantitatively 
consistent results, further demonstrating the robustness of our method. The
images in the three filters are used jointly to estimate the background 
galaxies' redshift distribution and so provide a correct normalization of 
the mass determination.

 We apply this method to the well-known cluster Abell~1689 as a test-case.
We find only one significant mass peak in the mass reconstructions, 
corresponding to the cluster itself. This is consistent with preliminary 
results from a large spectroscopic survey of Abell~1689 and its outskirts 
\citep{czoske04}, which shows that the environment of this cluster is 
remarkably smooth and quiet. We also compare our results to previous work 
by \citet{clowe01} who used an independent data set and the methods from 
\citet{kaiser95} and \citet{Kaiser-Squires1993} for their galaxy shape 
measurements and mass reconstruction. Within the errors both 
reconstructions agree very well. The same is true for the $M/L$ 
determination, which is consistent with previous findings. Moreover we are 
able to build a $M/L$ profile which in the case of Abell~1689 shows a near 
constant behaviour at large radius with a possible decrease close to the 
center. This suggests that mass traces light at least in the outskirts of 
the cluster. The drop of $M/L$ in the cluster centre may be due to an
underestimate of the mass in the centre, due to increasing contamination of
the background galaxy catalogue by cluster members diluting the lensing
signal. The flat $M/L$ profile in the infall region of the cluster 
indicates that the association between mass and light has already been 
achieved outside the cluster and the effect of the cluster environment on 
the mass-to-light ratio of infalling galaxies and groups is minor. This
supports the picture of a hierarchical assembly of clusters.

 For the results presented here we did not make use of the colour 
information available from multi-band imaging to separate cluster from
background galaxies which makes our results directly comparable to those of
\citet{clowe01}. However \citet{clowe03} presented an updated mass 
reconstruction for Abell~1689, this time using colours derived from our 
CFH12k images. The colour information resulted in an improved removal of 
cluster galaxies from his background galaxy catalogue, increasing both 
$r_{200}$ and $c$ for his best-fit NFW model and better agreement of the 
weak lensing mass profile with that derived from strong lensing. We will 
include colour selection of the different galaxy catalogues in a 
forthcoming paper aimed at comparing in great detail all the mass estimates
at different scales in Abell~1689: velocity distribution of the galaxies, 
X-ray mass maps, strong lensing in the center of the cluster and weak
lensing at larger scales. Provided the dynamics of the cluster is well
understood this should give a consistent picture of its mass distribution 
and components. This is the main goal of the pan-chromatic survey which is 
conducted by our group on intermediate redshift X-ray clusters.

 Finally, we will present a global study of our results based on the
application of the present methodology to the whole cluster catalogue, with
a discussion of the statistical properties of such clusters. A better 
understanding of the global properties of the mass distribution in rich 
clusters of galaxies will provide profound insights into the growth of 
structure in the Universe.

\begin{acknowledgements}
  We wish to thank Sarah Bridle and Phil Marshall regarding the many
  interaction and helpful discussions we had, specially regarding
  \textsc{Im2shape} and {\sc LensEnt2}. We wish to thank CALMIP
  (\emph{CALcul en MIdi-Pyr\'en\'ees}) for their data-processing resources
  during the last 2002 quarter, as the software used here is CPU-time and
  RAM consuming, and the Programme National de Cosmologie of the CNRS for
  financial support. JPK acknowledges support from CNRS and Caltech.
  IRS acknowledges support from the Royal Society.
\end{acknowledgements}

\renewcommand{\refname}{References} \bibliographystyle{aa}

\clearpage

\end{document}